

\long\def\UN#1{$\underline{{\vphantom{\hbox{#1}}}\smash{\hbox{#1}}}$}
\def\NP{\vfil\eject}
\def\NI{\noindent}

\magnification=\magstep 1
\overfullrule=0pt
\hfuzz=16pt
\voffset=0.0 true in
\vsize=8.8 true in
\baselineskip 20pt
\parskip 6pt
\hoffset=0.1 true in
\hsize=6.3 true in
\nopagenumbers
\pageno=1
\footline={\hfil -- {\folio} -- \hfil}

\hphantom{A}

\hphantom{A}

\centerline{\UN{\bf Exact Results for 1D Conserved Order Parameter
Model}}

\vskip 0.4in

\centerline{\bf Vladimir Privman}

\vskip 0.2in

\NI{\sl Department of Physics, Clarkson University,
Potsdam, New York 13699--5820, USA}

\vskip 0.4in

\centerline{\bf ABSTRACT}

Recent exact results for a particle-exchange model on a linear
lattice, with only irreversible moves reducing the local energy
allowed, are reviewed. This model describes a zero-temperature
Kawasaki-type phase separation process which reaches a frozen,
initial-condition-dependent state at large times.

\vfill

\NI {\bf PACS numbers:}\  82.20.$-$w,\  05.40.$+$j

\

\NP

This short review covers selected recent results of interest in
phase coarsening, spinodal decomposition, nucleation.$^{1-3}$
These processes can be modeled quite successfully by irreversible,
zero-temperature, low dimensional lattice
systems.$^{4-6}$  Variants of nonconserved order parameter dynamical
models in 1D, with effectively $T=0$ Glauber-type
spin dynamics, have been solved exactly for the
structure factor and average domain size, as functions of
time.$^{6-7}$ The mechanism leading to cluster growth in
1D is pairwise annihilation of interfaces separating ordered
domains; the interfacial motion is simple diffusion.

Spin-exchange, Kawasaki-type models, no longer allow for a simple,
diffusional interpretation of the dynamics. Derivation of exact
result even in 1D is more difficult than for the nonconserved
models; few very recent results were obtained, in 1D and on the
Bethe lattice, for a model with maximal limitation of the dynamical
moves allowed.$^{8-11}$ These new results will be surveyed here
and illustrated by working though the steps of a 1D
solution.$^{8,11}$ Somewhat different nomenclature and several
generalizations can be found in Ref.~10. The 1D solutions can also
be obtained by a different, cluster-size-distribution
approach$^{12}$ based on the low-$T$ Ising-model studies of
Ref.~13.

Detailed Bethe-lattice results were obtained in Ref.~9; these will
not be described here. The method of solution both in 1D and on
Bethe lattices bears similarity to the techniques employed in
kinetics of reactions with immobile reactants$^{14-15}$ and in
random sequential adsorption reviewed, e.g, in Refs.~16--17.

Let us emphasize that for \UN{nonconserved}
models, ordering in 1D in the Glauber-type dynamics
involves interface annihilation which is a process lowering the
local energy and therefore has Boltzmann factor $+\infty$
associated with its transition probability at $T = 0$. Interface
diffusion does not change the local energy and therefore has
Boltzmann factor $1$. Finally, interface generation (birth) has
Boltzmann factor $0$ (due to energy cost) at $T=0$. The $T=0$
models$^{6-7}$ with unlimited domain growth, $\sim \sqrt{t}$,
for large times $t$, correspond to allowing for both
annihilation and diffusion. Models with annihilation only lead to
frozen states.$^{14-15}$

For \UN{conserved}, spin-exchange, Kawasaki-type
models, ordering processes even at $T=0$ are more complicated than
for the nonconserved case. Specifically, let us consider the
binary AB-mixture model: each site of the lattice is occupied
by particle A or particle B. The dynamics generally involves nearby
particle exchanges; the locally conserved order parameter is the
difference of the A-\ and B-particle densities. If one only
eliminates energy-increasing moves, one is still left with a model
which was not solved exactly. Several numerical studies and
analytical expansion results were reported$^{18-22}$ for such
particle-exchange models in D up to 5. As in the nonconserved
case, some of the properties of the 1D models are different from
${\rm D} >1$. However, the general expectation of the ``freezing''
of the domain structure at large times applies, for conserved $T=0$
dynamics, in all space dimensions studied.

We now turn to the strictly 1D case. Let us assume that initially
the linear lattice sites are occupied by particles $A$ and
$B$ in exact alternating arrangement. We will solve the dynamics
of this model$^8$ under the assumption that unlike particle
contacts cost energy. The only dynamical moves allowed will be
those nearest-neighbor exchanges which reduce the energy, i.e., $A
\rightleftharpoons B$ exchanges in the configurations

$$\ldots ABAB \ldots \to \ldots AABB \ldots \;\; , \eqno(1)$$

$$\ldots BABA \ldots \to \ldots BBAA \ldots \;\; . \eqno(2)$$

\NI Indeed, these exchanges, of $AB$ or $BA$ pairs which are part
of a fully alternating sequence of 4 particles, reduce the local
number of unlike contacts from 3 to 1. The exchanges are
irreversible.

A model with this dynamics can be also solved for random initial
placement of particles,$^8$ with arbitrary initial density of $A$
and $B$ species,$^{11}$ which must fully cover the lattice. In all
cases, as the time goes by the number of ``reactive'' regions
decreases. For large times, a frozen state is approached at an
exponential rate.$^{8,11}$ Particle exchanges do not affect the
species concentrations. Therefore the dynamical quantity of
interest if the density of ``interfaces,'' i.e., of unlike
particle contacts, in the system, $I(t)$. We will present the
solution for the initially fully alternating state with $I(0)=1$.

We assume asynchronous, continuous-time dynamics. Each allowed
exchange event proceeds with the rate $R$ per unit time (which will
be absorbed in the dimensionless time variable, $Rt \to t$), and
independently of other exchange events. The method of solution for
various initial conditions, and for several related
models,$^{8-11,15}$ involves consideration of probabilities $P_k
(t) $ that a randomly selected continuous group of $k$ lattice
sites is fully ``reactive.''\ In the case chosen for illustration
here, there is the $A \rightleftharpoons B$ symmetry, while fully
reactive here means, for $k \geq 4$, fully alternating. (We will
also have to use the fully-alternating probability for 3
consecutive sites, $P_3(t)$, although ``reactivity'' cannot be
decided from 3 sites only.) In other applications, one considers
different objects, and no $A \rightleftharpoons B$ symmetry in
some instances.

Thus, $P_k (t)$ here is defined as the probability that a randomly
chosen, consecutive $k$-site sequence of sites is, at time
$t$, occupied by particles in the configurations $ABAB\ldots $ or
$BABA\ldots$. Note that the density of interfaces is then obtained
from the relation

$$ {dI(t) \over dt} = -2P_4 (t) \;\; , \eqno(3)$$

\NI which is self-explanatory in view of (1)-(2). The factor 2 is
due to the reduction from 3 to 1 of the number of interfaces in
each allowed particle exchange event.

The key to solubility is in that for some
dynamical rules (where details of course depend on the
model), the probabilities $P_k (t)$ are
determined locally, from the configuration of the clusters which
include the $k$-cluster. Furthermore, \UN{in one dimension} the
``topology'' of the surrounding clusters is trivial. This allows
to write a closed hierarchy of rate equations; in our case it is

$$ - {dP_k(t)  \over dt} = (k-3) P_k (t) + 2 P_{k+1} (t) +2 P_{k+2}
(t) \;\; . \eqno(4)$$

\NI This recursion must be considered for $k \geq 3$ and with the
initial condition

$$ P_k (0) = 1 \;\; , \eqno(5)$$

\NI because in the selected example of the fully alternating
initial configuration, all the $k$-site clusters are fully reactive
initially.

The rate equations (4) incorporate exchange processes with all the
``deciding'' four sites, see (1)-(2), within the original
$k$-cluster. There are $k-3$ such 4-site groups in the $k$-site
cluster. In addition, up to two exchange events may be possible
with the ``deciding'' 4-site group extending outside the original
cluster even though the actually exchanging $A$ and $B$ particles
are within it, at the end. The ``decision making'' is then
within the larger, $(k+1)$-clusters which include the original
$k$-cluster. The second term on the right-hand side of (4) accounts
for these events. Finally, up to two end-particles in the
$k$-cluster can exchange with nearby particles outside it. It is
obvious that the ``decision'' is then based on the configuration of
the two appropriate $(k+2)$-site clusters. The third term in (4)
accounts for such external exchanges.

Typically, 1D hierarchies such as (4) can be solved by assuming a
form exponential in $k$,

$$ P_k (t) = P_k (0) C(t) E^k(t) \;\; . \eqno(6)$$

\NI In our example, we find$^8$

$$ C(t)= \exp \left( 2e^{-t} + e^{-2t} -3 + 3t \right) \;\; ,
\eqno(7)$$

$$ E(t)=e^{-t} \;\; , \eqno(8)$$

\NI which can be checked by substitution in (4).

For the density of interfaces, we get, via (3), the result

$$ I(t)=1-2e^{-4} \int\limits_{1+e^{-t}}^2 e^{z^2} dz \;\; .
\eqno(9)$$

\NI Specifically, for large times the ordering process stops at
the residual density of unlike-particle contacts $I(\infty)
\simeq 0.450898$. More generally, both the saturation value and
time-variation are initial-condition dependent;$^{8,11}$
explicit results can be found in Refs.~8--11, 15.

In summary, we surveyed an approach to deriving exact
time-dependent solutions for 1D models with $T=0$ Kawasaki
dynamical moves restricted to the extreme irreversibility and lack
of ergodicity. The 1D results are useful as test models for
approximation schemes and numerical studies in D$>1$. Indeed,
standard approximation schemes developed for random
sequential adsorption, for instance (reviewed in Refs.~15--16), have
been based on hierarchies of equations similar to those that appear
in irreversible Kawasaki dynamics considered here. In D$>1$, they
involve complicated cluster shapes and must be truncated within
closure schemes. However, 1D exactly solvable hierarchies have
provided a useful guide to these approximation schemes.

\NP

\NI {\bf REFERENCES}{\frenchspacing

\item{1.} J.D.~Gunton, M.~San~Miguel, P.S.~Sahni,
{\sl Phase Transitions and Critical
Phenomena}, Vol.~8, p.~267, C.~Domb and J.L.~Lebowitz, eds.
(Academic, London, 1983).

\item{2.} A.~Sadiq and K.~Binder, J.~Stat. Phys. {\bf 35}, 517
(1984).

\item{3.} O.G.~Mouritsen, in {\sl Kinetics and
Ordering and Growth at Surfaces}, p.~1, M.G. Lagally, ed. (Plenum,
NY, 1990).

\item{4.} M. Scheucher and H. Spohn, J. Stat. Phys. {\bf 53}, 279
(1988).

\item{5.} B. Hede and V. Privman, J. Stat. Phys. {\bf 65}, 379
(1991).

\item{6.} V. Privman, J. Stat. Phys. {\bf 69}, 629 (1992).

\item{7.} A.J. Bray, J. Phys. A{\bf 23}, L67 (1990).

\item{8.} V. Privman, Phys. Rev. Lett. {\bf 69}, 3686
(1992).

\item{9.} S.N. Majumdar and C. Sire, Phys. Rev. Lett. {\bf 70},
4022 (1993).

\item{10.} P.L. Krapivsky, J. Stat. Phys., in print.

\item{11.} J.--C. Lin and P.L. Taylor, preprint.

\item{12.} S.J. Cornell, preprint.

\item{13.} S.J. Cornell, K. Kaski and R.B. Stinchcombe, Phys. Rev.
B{\bf 44}, 12263 (1991).

\item{14.} V.M. Kenkre and H.M. Van Horn, Phys. Rev.
A{\bf 23}, 3200 (1981).

\item{15.} S.N. Majumdar and V. Privman, J. Phys. A{\bf
26}, L743 (1993).

\item{16.} M.C.~Bartelt and V.~Privman, Int.
J.~Mod. Phys. B{\bf 5}, 2883 (1991).

\item{17.} J.W.~Evans, Rev. Mod. Phys., in print.

\item{18.} A. Levy, S. Reich and P. Meakin, Phys. Lett. {\bf 87}A,
248 (1982).

\item{19.} P. Meakin and S. Reich, Phys. Lett. {\bf 92}A, 247
(1982).

\item{20.} R.G. Palmer and H.L. Frisch, J. Stat. Phys. {\bf 38},
867 (1985).

\item{21.} F.C. Alcaraz, J.R. Drugowich de Fel\'icio and R.
K\"oberle, Phys. Lett. {\bf 118}A, 200 (1986).

\item{22.} Y. Elskens and H.L. Frisch, J. Stat. Phys. {\bf 48},
1243 (1987).

}\bye